\documentclass[conference]{IEEEtran}
\IEEEoverridecommandlockouts
\usepackage[tight,footnotesize]{subfigure}
\usepackage{float}
\usepackage{hyperref}
\usepackage{bm}
\usepackage{cite}
\usepackage{graphicx}
\usepackage[ruled]{algorithm2e} 
\usepackage{algorithmic}
\usepackage{amssymb}
\usepackage{amsmath}
\usepackage{amsthm}
\usepackage{mathrsfs}
\usepackage{color}
\usepackage{booktabs}
\usepackage{framed}
\usepackage{adjustbox}

\theoremstyle{definition}

\allowdisplaybreaks[1]
\allowdisplaybreaks

\def\BibTeX{{\rm B\kern-.05em{\sc i\kern-.025em b}\kern-.08em
    T\kern-.1667em\lower.7ex\hbox{E}\kern-.125emX}}
    
\begin{document}

\title{ 
 Communication-Efficient Cooperative Localization: A Graph Neural Network Approach
}

\author{Yinan Zou, Christopher G. Brinton, and  Vishrant Tripathi\\ Elmore Family School of Electrical and Computer Engineering\\
Purdue University, West Lafayette, IN 47907, USA}

\vspace{-5mm}

\maketitle
\begin{abstract}
Cooperative localization leverages noisy inter-node distance measurements and exchanged wireless messages to estimate node positions in a wireless network. 
In communication-constrained environments, however, transmitting large messages becomes problematic. 
In this paper, we propose an approach for communication-efficient cooperative localization 
that addresses two main challenges. First, cooperative localization often needs to be performed over wireless networks with loopy graph topologies. Second is the need for designing an algorithm that has low localization error while simultaneously requiring a much lower communication overhead.
Existing methods fall short of addressing these
two challenges concurrently. To achieve this, we propose a vector quantized message passing neural network (VQ-MPNN) for cooperative localization.
Through end-to-end neural network training, VQ-MPNN enables the co-design of node localization and message compression.
Specifically, VQ-MPNN treats prior node positions and distance measurements as node and edge features, respectively, which are encoded as node and edge states using a graph neural network.
To find an efficient representation for the node state, we construct a vector quantized codebook for all node states 
such that instead of sending long messages, each node only needs to transmit a codeword index. 
Numerical evaluations demonstrates that our proposed VQ-MPNN approach can deliver localization errors that are similar to existing approaches while reducing the overall communication overhead by an order of magnitude.

\end{abstract}

\section{Introduction}

Localization of node positions is an essential functionality for many wireless networks. In harsh
environments, such as underwater networks, caves, and dense forests, satellite-based positioning systems such as GPS/GNSS are often inaccessible, requiring alternative methods for position estimation. In such settings, there are usually a few anchor nodes with known fixed positions and numerous agent nodes with unknown positions. To utilize this structure,  
\textit{cooperative localization} methods \cite{patwari2005locating,xiong2021cooperative} have been proposed, where agent nodes localize themselves using inter-node measurements and exchanged messages.
Cooperative localization is more flexibile, scalable, and robust than centralized implementations and plays a crucial role in modern sensing systems.

\subsection{Cooperative Localization: Overview and Challenges}
Cooperative localization typically involves two phases:
a distance measurement phase and a location-update phase.
First, during the distance measurement phase, each node measures the inter-node distance using metrics such as received signal strength (RSS), time of arrival (TOA), or round-trip time of arrival (RTOA).
Then, during the location-update phase, each node exchanges messages with distance measurements and estimated positions over multiple iterations, eventually using all the information to infer its own position.

\begin{figure}[tbp]
	\centering
\includegraphics[width=0.8\linewidth]{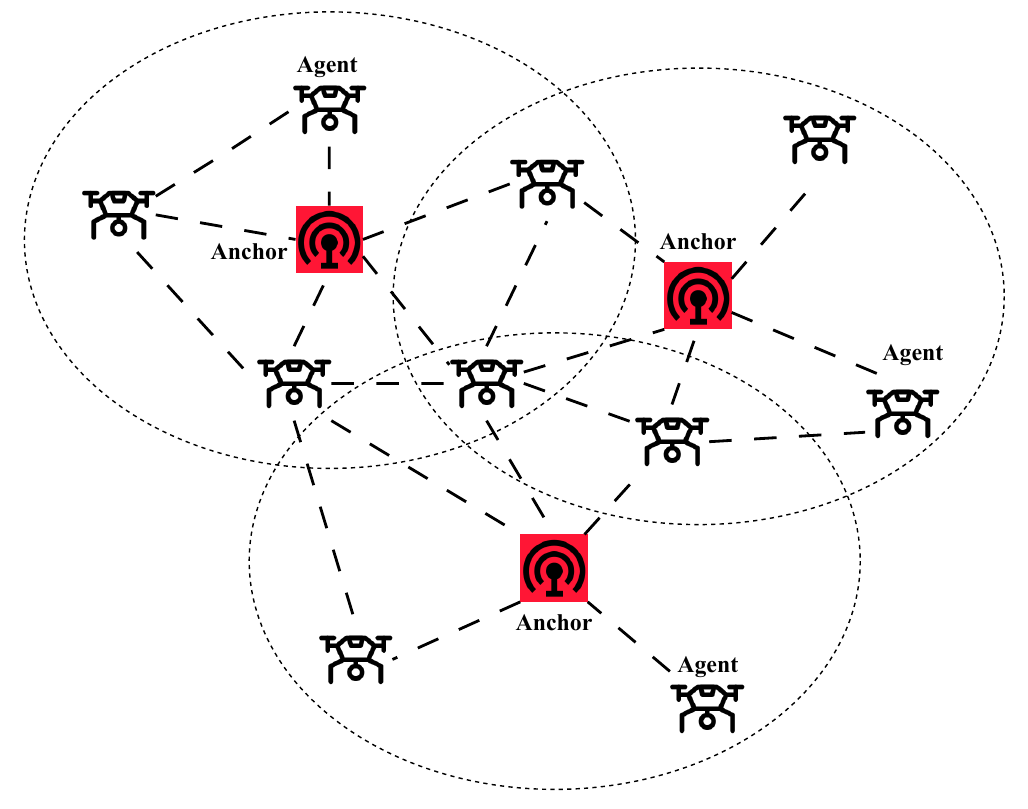}
	\caption{
    The cooperative localization framework consists of a few anchors with fixed known locations and many agents with unknown locations. Both anchors and agents communicate with neighboring nodes to measure relative distances and then exchange messages iteratively to perform global localization.
	} 
	\label{framework}
    \vspace{-6mm}
\end{figure}
Belief propagation (BP) \cite{wymeersch2009cooperative,ihler2005nonparametric,meyer2018message}
and distributed optimization \cite{piovesan2016cooperative,gur2020alternating,zhang2023distributed}  are two widely used approaches for the location-update phase.
The distributed optimization methods are non-Bayesian and treat agents' positions as deterministic parameters while disregarding prior information.
In contrast, BP, as a Bayesian method, treats agents' positions as random variables, which allows for the quantification of estimation uncertainties.
BP can be categorized into two types based on the representation of messages and beliefs: parametric BP and nonparametric BP. 
Parametric BP typically employs the mean and covariance of Gaussian distributions to represent messages and beliefs. 
In contrast, nonparametric BP employs thousands of weighted samples to depict messages in detail, enabling the representation of distributions with arbitrary shapes.
For tree-structured graphs, BP works well and can provide the precise marginal posterior.
However, physical graphs existing in cooperative localization networks often include many loops.
For loopy graphs,
BP can only obtain an approximation of the true marginal posterior,
leading to suboptimal
estimates
\cite{li2022convergence,weiss1999correctness,yedidia2005constructing,riegler2012merging,satorras2021neural}.
Existing BP-based methods employ approximation techniques to relax the original problem into a more tractable form, which typically yields suboptimal solutions.

During the location-update phase, agent nodes need to exchange message with each other over multiple iterations. 
	In nonparametric BP, iterative transmission of messages represented by weighted particles often encounters a fundamental communication bottleneck due to limited spectrum resources\cite{asadi2014survey}, 
    necessitating efforts to reduce communication overhead.
    Compression techniques, such as quantization \cite{ihler2005loopy}, Gaussian mixture models \cite{ihler2005nonparametric,savic2012reducing}, and partial
    mutual information \cite{mendrzik2016particle}, can be used to compress messages in nonparametric BP, which however introduce unavoidable  compression errors. 
    
Compressing messages to reduce communication overhead while maintaining localization performance thus remains an open problem. To address this question, in this work, we investigate data-driven techniques. In particular, graph neural networks (GNNs) have recently been utilized for graph-related tasks to rectify errors in loopy belief propagation \cite{
satorras2021neural,adiga2024generalization,
liang2021neural, tedeschini2023message,cao2024distributed}.
By exchanging information between nodes, GNNs effectively capture topological dependencies.
Given that neural networks excel at function approximation, we hypothesize that GNNs can be employed to learn the mapping from distance measurements to actual node positions, which has the potential to improve cooperative localization performance.


\vspace{-0.05in}
\subsection{Our Approach: Overview and Contributions}
In this paper, we study communication-efficient cooperative localization problem where agent nodes localize themselves using inter-node measurements and wireless messages exchanged under communication constraints. We address two main challenges.
First, cooperative localization often needs to be performed over wireless networks with loopy graph topologies. Second, we need to design an algorithm that has low localization error while simultaneously requiring a much lower communication overhead.
Existing methods fall short of concurrently addressing these two challenges. 
To simultaneously deal with these two challenges, we propose a vector quantized message passing neural network (VQ-MPNN).
Through  end-to-end neural network training, VQ-MPNN enables the co-design of node localization and message compression. Our approach can handle loopy graphs (unlike belief propagation based methods) while having a communication overhead that is an order of magnitude less than the best known deep learning based methods. 
A primary distinction between our proposed VQ-MPNN and MPNN approaches in existing works \cite{liang2021neural, tedeschini2023message, cao2024distributed} lies in the construction of a compact VQ codebook for message representation, which reduces the communication cost to just a few bits per message.

Our main contributions are summarized as follows:

\begin{itemize}
\item 
We formulate VQ-MPNN to treat prior node positions and distance measurements as node and edge features, respectively, and design its graph neural network to encode these features as node and edge states.
During each communication round, then, nodes engage in cooperative message-passing to update their node and edge states. After several communication rounds, the final node state can be fed into the estimation neural network to compute the estimated node position.

\item To find an efficient representation for the node state, we construct a vector quantized (VQ) codebook for all node states. The VQ codebook is regarded as a trainable parameter and is constructed during the neural network training phase. Each node only needs to transmit the codeword index rather than the node state, which we find \textit{reduces communication cost by an order of magnitude}.

\item Benefiting from the message passing mechanism, the proposed VQ-MPNN approach can perform inference on the graph corresponding to the network topology in a distributed manner. Importantly, even if nodes are added or removed from
the graph, the proposed neural network still works, which
demonstrates its strong generalizability and scalability.

\item Numerical evaluations demonstrate that our proposed approach outperforms or matches existing solutions like PLBP \cite{garcia2017cooperative} and SP-ADMM \cite{zhang2023distributed} in localization accuracy, while drastically improving communication efficiency.
\end{itemize}

 \begin{figure*}[tbp]
	\centering
	\includegraphics[width=0.9\linewidth]{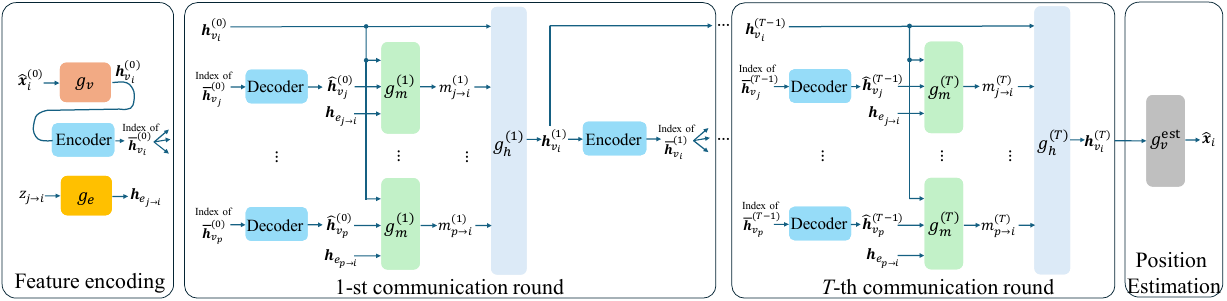}
	\caption{VQ-MPNN architecture. 
   Each node $i$ takes its prior position and distance measurements as inputs, encoding them into node and edge states.
    Then, each node applies the encoder to map its node state to a codeword index,  which is transmitted to neighboring nodes.
    During each communication round, node $i$ receives the codeword indices from its neighbors and reconstructs them into node states. 
    Each node then integrates these reconstructed node states with its own node and edge states to update its node state. After several communication rounds, each node uses its final updated node state to estimate its position.} 
	\label{VQ_MPNN_archi}
\vspace{-6mm}
\end{figure*}

\vspace{-0.05in}
\subsection{Related Work}
In recent years, numerous studies have focused on cooperative localization.
Given the distributed nature of computation, message passing algorithms based on probabilistic graphical models \cite{wymeersch2009cooperative,ihler2005nonparametric,meyer2018message} and 
distributed optimization methods \cite{zhang2023distributed,piovesan2016cooperative,gur2020alternating} have been utilized for cooperative localization in prior works.
In the distributed methods, each node localizes itself using local network information from its neighbors rather than relying on data from the entire network, making these methods well-suited for large-scale networks.

In the Bayesian framework, BP can be employed to obtain the marginal posterior distribution of a node's position.
BP runs on the factor graph to calculate the marginal posteriors by passing messages between nodes.
Since the distance measurement model is usually non-linear, parametric BP approximates the non-linear model using Taylor expansions \cite{li2022convergence,li2015gaussian}, statistical linear regression \cite{garcia2017cooperative}, or variational posterior approximation \cite{yu2024robust}.
Typically, all the BP messages and
beliefs are derived in Gaussian form, with only their means and covariances being updated and exchanged during the process. However, the approximation of nonlinear models via linear ones leads to model mismatch and accuracy issues, thereby degrading localization performance.

For nonparametric BP, the authors in \cite{ihler2005nonparametric,lien2012comparison,kim2018connectivity}  use weighted particles to represent messages and beliefs, enabling the flexible representation of distributions in any shape.
The iterative
broadcasting and aggregation of particles result in significant
communication costs.
To reduce the communication cost of particle transmission,
the authors in \cite{mendrzik2016particle} proposes selecting important particles for transmission based on partial mutual information.
The authors in \cite{savic2012reducing,ihler2005nonparametric} utilize the Gaussian
mixture model to approximate the particles, enabling the
exchange of only the mixture representation parameters among
agents.
Although the aforementioned works utilize compression techniques to minimize communication cost, they inevitably introduce compression errors. Further, both parametric and nonparametric BP methods struggle to perform well on topologies with loops.

Recently, graph neural networks (GNNs) have demonstrated excellent performance in various graph-related tasks, including cooperative localization.
The authors in \cite{liang2021neural,cao2024distributed} introduce the message passing neural network (MPNN) into nonparametric and parametric BP to address the cooperative localization problem modeled by loopy factor graph. 
By considering the temporal correlation in the motion
model of agents, \cite{tedeschini2023message}  combines a long short-term memory
(LSTM) network with a MPNN for agent tracking.
Motivated by these works, we propose a GNN based approach that can simultaneously handle loopy graphs and have much lower communication overhead.

\section{System Model}

Consider a wireless network consisting of $N_s$  
anchor nodes with precisely known positions and $N_a$ agent nodes whose positions are unknown.
The index sets of anchor and agent nodes are denoted as $\mathcal{I}_s$ and $\mathcal{I}_a$.
The total set of nodes $\mathcal{I}$ is defined as $\mathcal{I} = \mathcal{I}_s\cup\mathcal{I}_a$.
We denote $\bm{x}_i\in \mathbb{R}^2$ as the two dimensional position of node $i, \forall i \in 
\mathcal{I}$.
We assume a known prior distribution on positions given by $p(\bm{x}_i)$.
Each agent node receives noisy distance measurements to neighboring nodes within a communication range $r$ and communicates only with these neighboring nodes.
It is assumed that each node is aware of its neighboring nodes. 
Fig. \ref{framework} illustrates the cooperative localization framework.

Cooperative localization comprises of two phases: the distance measurement phase and location update phase. 
In the distance measurement phase, each agent node obtains noisy distance measurement from the received pilot signals of neighboring nodes.
Measurements can be obtained through various methods, with the most common approaches including RSS, TOA, or RTOA.  
When node $j \in \mathcal{N}_i $ transmits a  pilot signal to node $i \in \mathcal{I}_a$,
the distance measurement from node $j$ to $i$ is modeled as
\begin{align}
	{z}_{j\to i} = d(\bm{x}_j,\bm{x}_i) + {n}_{j\to i},\quad j \in \mathcal{N}_i,
\end{align}
where $d(\bm{x}_j,\bm{x}_i) = \|\bm{x}_j-\bm{x}_i\|_2$ denotes true distance between node $i$ and $j$,
$\mathcal{N}_i \subseteq \mathcal{I}$ denotes the neighbors of node $i$, and ${n}_{j\to i}$ denotes measurement noise.
Only node $i$ has access to the measurement ${z}_{j\to i}$.
Despite the fact that $d(\bm{x}_j,\bm{x}_i) = d(\bm{x}_i,\bm{x}_j)$,
the measurement ${z}_{j\to i}$ is not necessarily equal to ${z}_{i\to j}$ because ${n}_{j\to i}$ is not necessarily equal to ${n}_{i\to j}$.
We assume that each node has already obtained distance measurements from its neighbors. 

In the location-update phase, 
each node exchanges messages 
containing distance measurements and estimated positions with neighboring nodes over multiple iterations, and then uses all the collected information to infer its position.
Since we do not have full network connectivity, we need multiple iterations of message exchanges to obtain all the information needed to infer global node positions accurately.

\section{Communication-Efficient Graph Neural Network: VQ-MPNN}

In this section, we propose VQ-MPNN for cooperative localization.
Fig. \ref{VQ_MPNN_archi} shows the overall architecture.

\subsection{Graph Representation}
We characterize the cooperative localization network as a
directed graph, where the agent and anchor nodes are regarded
as graph nodes and the distance measurements between nodes
are regarded as edges.
At initialization, the initial position $\hat{\bm{x}}_i^{(0)}$ is drawn from prior distribution $p(\bm{x}_i)$.
Note that the node feature associated with node $i$ is the initial position $\hat{\bm{x}}_i^{(0)}$, while the edge feature associated with each edge from node $j$ to node $i$ is the distance measurement ${z}_{j\to i}$.

 \subsection{VQ-MPNN Architecture}

\begin{figure}[tbp]
	\centering
	\includegraphics[width=0.99\linewidth]{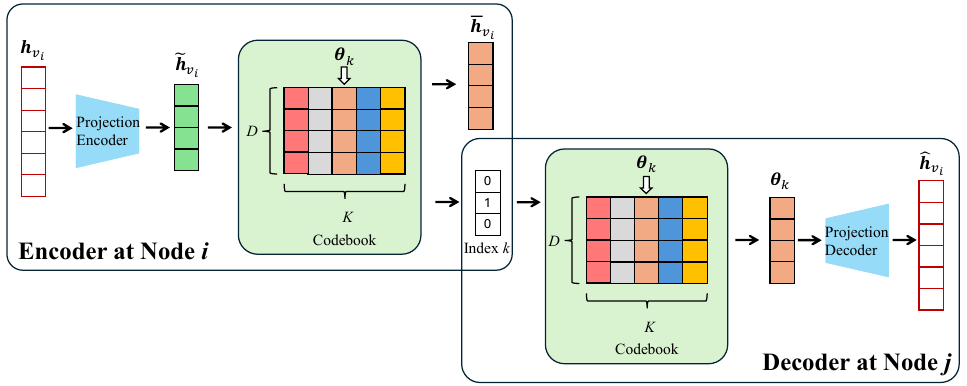}
	\caption{Encoder and decoder for node states.
    Node $i$ employs the projection encoder to map its node state into a latent vector, determines the nearest codeword to the latent vector, and transmits the corresponding codeword index to its neighbors. 
    Neighboring node $j$ receives the codeword index, retrieves the corresponding codeword from the codebook, and inputs this codeword into the projection decoder to recover the node state.
    } 
	\label{VQVAE_archi}
    \vspace{-8mm}
\end{figure}

	\begin{itemize}
	\item[(1)]
		\emph{Feature Encoding}:
		We utilize Multi-Layer Perceptrons (MLPs) 
  to map node and edge features into an $M$-dimensional unified space 
	\begin{align}
		&\bm{h}_{{v}_i}^{(0)} = g_v(\hat{\bm{x}}_i^{(0)}),  \label{eq23}
		\\&\bm{h}_{{{e}}_{j\to i}}	= g_e({z}_{j\to i}), \label{eq24}
	\end{align}
	where $g_v(\cdot):\mathbb{R}^2\to\mathbb{R}^M$ and $g_e(\cdot):\mathbb{R}^1\to\mathbb{R}^M$ are MLPs, 
 	$\bm{h}_{{v}_i}\in \mathbb{R}^M$ and $\bm{h}_{{{e}}_{j\to i}}\in \mathbb{R}^M$ denote the node state and the edge state, respectively.
    Subsequently, each node  uses an encoder to map the $M$-dimensional node state to a $D$-dimensional codeword within the VQ codebook, which contains $K$ codewords, and retrieves the corresponding codeword index.
		The encoder is given by
		\begin{align}
			\text{Index}(\bar{\bm{h}}_{{v}_i}^{(0)}) = \text{Encoder}(\bm{h}_{{v}_i}^{(0)}),
		\end{align}
		where $\text{Index}(\cdot)$ denotes the index of a codeword and $\bar{\bm{h}}_{{v}_i} \in \mathbb{R}^D$ denotes quantized node state.
        Note that the quantized node state is a codeword in the codebook. 
        The codebook index is represented by binary code.
        For instance, in a VQ codebook with $K=5$, the index of the third codeword $\theta_3$ is represented as 010, as illustrated in Fig. \ref{VQVAE_archi}.
		The design of the encoder will be elaborated in section \ref{vqvae_design}.
		
	\item[(2)] \emph{Inter-node Communication}:
        We denote the maximum number of communication rounds as $T$.
	In the communication round $t\in\{1,\ldots,T\}$, each node $i$  broadcasts the index of its quantized node state and receives the indices of quantized node states from its neighbors.
	This exchange of node states is analogous to the beliefs exchange in BP.
	Utilizing the index of the quantized node state, each node $i$ reconstructs the node state using a decoder as follows 
	\begin{align}
		\hat{\bm{h}}_{{v}_j}^{(t-1)} = \text{Decoder}(\text{Index}(\bar{\bm{h}}_{{v}_j}^{(t-1)})),
	\end{align}
	where $\hat{\bm{h}}_{{v}_j} \in \mathbb{R}^M$ denotes the recovered node state.
    The design of decoder will be elaborated in section \ref{vqvae_design}.
	Subsequently, we concatenate the edge state and the node states into a single vector and input this vector into a MLP to generate the message
	\begin{align}\label{eq09}
		\bm{m}_{j\to i}^{(t)} = g_m^{(t)}(  \hat{\bm{h}}_{{v}_j}^{(t-1)} , \bm{h}_{{v}_i}^{(t-1)},
		\bm{h}_{{{e}}_{j\to i}}),
	\end{align}
        where $\bm{m}_{j\to i} \in\mathbb{R}^M$ denotes the message, $g_m^{(t)}( \cdot):\mathbb{R}^{3M}\to\mathbb{R}^M$ denotes a MLP, $\hat{\bm{h}}_{{v}_j}$ denotes the recovered node states of node $j$, $\bm{h}_{{v}_i}$ denotes the node state of node $i$, and $\bm{h}_{{{e}}_{j\to i}}$ denotes the edge state of edge from node $j$ to node $i$.
		
	\item[(3)] \emph{Combination}: In the communication round $t\in\{1,\ldots,T\}$, each node $i$ integrates its previous node state with messages to obtain the updated node state
	\begin{align}\label{eq10}
		\bm{h}_{{v}_i}^{(t)} = g_h^{(t)}(\bm{h}_{{v}_i}^{(t-1)}, \sum(\{\bm{m}_{j\to i}^{(t)}\}_{j\in\mathcal{N}_i})),
	\end{align}
	where $\sum(\cdot)$ denotes the summation function, $g_h^{(t)}( \cdot):\mathbb{R}^{2M}\to\mathbb{R}^M$ denotes a MLP.
	Next, each node employs the encoder to generate the quantized node state for the subsequent communication round
	\begin{align}
		\text{Index}(\bar{\bm{h}}_{{v}_i}^{(t)}) = \text{Encoder}(\bm{h}_{{v}_i}^{(t)}).
	\end{align}
	\item[(4)] \emph{Position Estimation}: After $T$ communication rounds, we input the final node state into the estimation neural network to obtain the estimated node position
	\begin{align}\label{eq12}
		\hat{\bm{x}}_i = g_v^{\text{est}}(\bm{h}_{{v}_i}^{(T)}),
	\end{align}
    where $\hat{\bm{x}}_i\in\mathbb{R}^2$ denotes the estimated node position and $g_v^{\text{est}}( \cdot):\mathbb{R}^{M}\to\mathbb{R}^2$ denotes a MLP.

\end{itemize}
Alg. \ref{workflow-VQMPNN} outlines  the workflow of the proposed VQ-MPNN for communication-efficient cooperative localization.

\subsection{Encoder and Decoder Design}\label{vqvae_design}

Motivated by \cite{van2017neural}, we construct a VQ codebook for node states that is shared among all nodes.
In particular, the codebook is represented as 
$\mathcal{C}\in\mathbb{R}^{D\times K}$, where $D$ denotes the length of the codeword and $K$ denotes the number of the codewords.
In other words, 
there are $K$ codewords, each with $D$ dimensions, represented as 
 $\bm{\theta}_k\in \mathbb{R}^D, k\in\{1,\ldots,K\}$.
All of the codewords are trainable parameters and are optimized during the training phase.
 
The process of utilizing encoder to compress node states is outlined as follows.
For sake of simplicity, we omit the index of the communication round $t$. 
Specifically, we use a projection encoder to map the node state $\bm{h}_{v_i}\in\mathbb{R}^M$ to the latent vector $\tilde{\bm{h}}_{v_i}\in\mathbb{R}^D$ as follows 
\begin{align}
	\tilde{\bm{h}}_{v_i} = \text{ProjEncoder}(\bm{h}_{v_i}),
\end{align}
where $\text{ProjEncoder}(\cdot):\mathbb{R}^{M}\to\mathbb{R}^D$ is a MLP.
Next, the nearest codeword to the latent vector $\tilde{\bm{h}}_{v_i}$ is determined  as follows
\begin{align}
	k  = \arg\min_j \|\tilde{\bm{h}}_{v_i}-\bm{\theta}_j\|_2,
\end{align}
where $k$ denotes the index of the nearest codeword.
Thus, the quantized node state corresponds to the codeword $\bm{\theta}_k$ as follows
\begin{align}
	\bar{\bm{h}}_{v_i} = \bm{\theta}_k.
\end{align}
Through the encoding process, each node only needs to transmit the index $k$ of the codeword instead of the original high-dimensional node state.

Upon receiving the index $k$, each node locates the corresponding codeword in the codebook to obtain $\bm{\theta}_k$ and then recovers the node state using the projection decoder 
\begin{align}
	\hat{\bm{h}}_{{v}_i} = \text{ProjDecoder}(\bm{\theta}_k),
\end{align}
where $\text{ProjDecoder}(\cdot):\mathbb{R}^{D}\to\mathbb{R}^M$ is a MLP.

In prior works on using MPNN for cooperative localization \cite{tedeschini2023message,liang2021neural} node states are directly exchanged among nodes. 
In contrast, we construct a VQ codebook to represent all node states, requiring only codeword indices to be transmitted, 
which significantly reduces the communication overhead.

\vspace{-2mm}
\subsection{Communication Cost}
The communication cost per round per node is the number of bits required for transmitting message to neighboring nodes and  the fixed overhead for preambles or headers of message packets.
The fixed overhead is assumed to be $H$ bits per message.
In the proposed VQ-MPNN, the communication bits for transmitting the index of the codeword is $\lceil\log_2 K  \rceil$.
Thus, VQ-MPNN has a total communication cost of $( H+\lceil\log_2 K  \rceil) N_i T $ bits per node.

\begin{algorithm}[tbp]
	\caption{Workflow of the proposed VQ-MPNN}
	\begin{algorithmic}[1]
		\STATE \textbf{Input:} Initial position $\hat{\bm{x}}_i^{(0)}$ and distance measurements $\{{z}_{j\to i}\}_{j\in\mathcal{N}_i}$.
		\STATE \textbf{Output:} 
        Estimated node position 
        $\hat{\bm{x}}_i$.
		\FOR {each node $i$}
        \STATE $\bm{h}_{{v}_i}^{(0)} = g_v(\hat{\bm{x}}_i^{(0)}),$
           $\bm{h}_{{{e}}_{j\to i}}	= g_e({z}_{j\to i}),\,\forall j \in\mathcal{N}_i$
        \STATE $\text{Index}(\bar{\bm{h}}_{{v}_i}^{(0)}) = \text{Encoder}(\bm{h}_{{v}_i}^{(0)})$
        \ENDFOR
		\FOR {communicaton round $t\in \{1,\ldots,T\}$}
		\FOR {each node $i$}
        \STATE Broadcast $\text{Index}(\bar{\bm{h}}_{{v}_i}^{(t-1)})$
		\STATE Receive
        $\{\text{Index}(\bar{\bm{h}}_{{v}_j}^{(t-1)})\}_{j\in\mathcal{N}_i}$ from its neighbors
        \STATE Recover node states:
        $\hat{\bm{h}}_{{v}_j}^{(t-1)} = \text{Decoder}(\text{Index}(\bar{\bm{h}}_{{v}_j}^{(t-1)})),\,\forall j \in\mathcal{N}_i$
        \STATE 
        Construct Message:
        $\bm{m}_{j\to i}^{(t)}= g_{m}^{(t)}(\hat{\bm{h}}_{{v}_j}^{(t-1)}, \bm{h}_{{v}_i}^{(t-1)},\bm{h}_{{{e}}_{j\to i}}),\,\forall j \in\mathcal{N}_i$
		\STATE Update node state:	
		$
		\bm{h}_{{v}_i}^{(t)} = g_h^{(t)}(\bm{h}_{{v}_i}^{(t-1)}, \sum(\{\bm{m}_{j\to i}^{(t)}\}_{j\in\mathcal{N}_i}))
		$
        \STATE
        Obtain codeword index: $\text{Index}(\bar{\bm{h}}_{{v}_i}^{(t)})
		 = \text{Encoder}(\bm{h}_{{v}_i}^{(t)})$
		\STATE Broadcast $\text{Index}(\bar{\bm{h}}_{{v}_i}^{(t)})$
		\ENDFOR
		\ENDFOR
    \FOR {each node $i$}
    \STATE
    $\hat{\bm{x}}_i = g_v^{\text{est}}(\bm{h}_{{v}_i}^{(\text{T})})$
    \ENDFOR
	\end{algorithmic}
    \label{workflow-VQMPNN}
\end{algorithm}

\begin{table*}[htb]
	\centering	\makeatletter\def\@captype{table}\makeatother\caption{Neural Network Setting}
	\begin{adjustbox}{width=0.97\textwidth}
	\begin{tabular}{|cc|cc|cc|cc|cc|cc|cc|cc|}
		\hline
		\multicolumn{2}{|c|}{$g_v(\cdot)$} 
		& \multicolumn{2}{c|}{$g_e(\cdot)$} 
		& \multicolumn{2}{c|}{$g_m^{(t)}(\cdot)$}
		& \multicolumn{2}{c|}{$g_h^{(t)}(\cdot)$}   
		& \multicolumn{2}{c|}{$g_v^{\text{est}}(\cdot)$}                 
		\\ \hline
		\multicolumn{1}{|c|}{Layer}   & Size 
		& \multicolumn{1}{c|}{Layer} & \multicolumn{1}{c|}{Size} 
		& \multicolumn{1}{c|}{Layer} & \multicolumn{1}{c|}{Size} 
		& \multicolumn{1}{c|}{Layer} & \multicolumn{1}{c|}{Size} 
		& \multicolumn{1}{c|}{Layer} & \multicolumn{1}{c|}{Size} 
		\\ \hline
		\multicolumn{1}{|c|}{Linear + GELU}    &    $2\times 64 $
		& \multicolumn{1}{c|}{Linear + GELU}   &    $1\times 32 $
		& \multicolumn{1}{c|}{Linear + GELU}   &    $3M\times 80 $
  & \multicolumn{1}{c|}{Linear + GELU}   &   $2M\times M $
		& \multicolumn{1}{c|}{Linear + GELU}   &   $M\times 128 $ 
		\\ 
		\multicolumn{1}{|c|}{Linear + GELU}  &     $64\times M $ 
		&\multicolumn{1}{c|}{Linear + GELU}  &     $32\times 64 $ 
		&\multicolumn{1}{c|}{Linear + GELU}  &     $80\times 16 $
		& \multicolumn{1}{c|}{} &   
		& \multicolumn{1}{c|}{Linear + GELU} &   $128\times 256 $
		\\
		\multicolumn{1}{|c|}{}  &     
		&\multicolumn{1}{c|}{Linear + GELU}  &     $64\times 32 $ 
		&\multicolumn{1}{c|}{Linear + GELU}  &     $16\times M $
		& \multicolumn{1}{c|}{} &     
		& \multicolumn{1}{c|}{Linear + GELU} &   $256\times 128 $
		\\
		\multicolumn{1}{|c|}{}  &     
		&\multicolumn{1}{c|}{Linear + GELU}  &     $32\times M$ 
		&\multicolumn{1}{c|}{}  &     
		& \multicolumn{1}{c|}{} &   
		& \multicolumn{1}{c|}{Linear} &   $128\times 2 $
		\\ \hline
	\end{tabular}
	\end{adjustbox}
	\label{VQMPNNsetting}
    \vspace{-2mm}
\end{table*}

\begin{table*}[htb]
	\centering
	\begin{tabular}{|cc|cc|cc|cc|cc|cc|cc|cc|}
		\hline
		\multicolumn{2}{|c|}{Projection Encoder} 
		& \multicolumn{2}{c|}{Codebook} 
		& \multicolumn{2}{c|}{Projection Decoder}              
		\\ \hline
		\multicolumn{1}{|c|}{Layer}   & Size 
		& \multicolumn{1}{c|}{Layer} & \multicolumn{1}{c|}{Size} 
		& \multicolumn{1}{c|}{Layer} & \multicolumn{1}{c|}{Size} 
		\\ \hline
		\multicolumn{1}{|c|}{Linear + GELU}    &    $M\times 16$  
		& \multicolumn{1}{c|}{Embedding}   &    $K\times D$ 
		& \multicolumn{1}{c|}{Linear + GELU}   &   $D\times 16$ 
		\\ 
		\multicolumn{1}{|c|}{Linear + GELU}  &   $16\times D$   &\multicolumn{1}{c|}{}  &       &\multicolumn{1}{c|}{Linear + GELU}  &   $16\times M$ 
		\\ \hline
	\end{tabular}
    \vspace{-2mm}
\end{table*}

\subsection{Training Strategy}
To train the proposed VQ-MPNN, we create a loss function that consists of 
the mean squared error (MSE) and VQ loss
\begin{align}
	&\mathcal{L}
	= \mathcal{L}_{\text{MSE}} + \mathcal{L}_{\text{VQ}}
	= \underbrace{
	\sum_{i=1}^{I}\|\hat{\bm{x}}_i - \bm{x}_i \|_2^2}_{\mathcal{L}_{\text{MSE}}}
	\notag
	\\&+
	\underbrace{
	\alpha \cdot	\bigg(
	\sum_{i=1}^{I}
	\sum_{i=1}^{T}
	\|\bm{h}_{v_i}^{(t)} - \text{ProjDecoder}(\tilde{\bm{h}}_{{v}_i}^{(t)} + \text{sg}(\bar{\bm{h}}_{{v}_j}^{(t)}-\tilde{\bm{h}}_{{v}_i}^{(t)}))\|_2^2} \notag
	\\&
	\underbrace{
	+ 
	\| \text{sg}(\tilde{\bm{h}}_{{v}_i}^{(t)}) - \bar{\bm{h}}_{{v}_j}^{(t)} \|_2^2
	+ \beta\cdot
	\| \tilde{\bm{h}}_{{v}_i}^{(t)} -\text{sg}( \bar{\bm{h}}_{{v}_j}^{(t)}) \|_2^2\bigg) }_{\mathcal{L}_{\text{VQ}}}.
\end{align}
Here $\alpha$ and $\beta$ denote penalty parameters, and $\text{sg}(\cdot)$ denotes the stopgradient operator \cite{van2017neural}.
We set $\alpha=0.1$ to make MSE loss and VQ loss are on the same order of magnitude.
We follow \cite{van2017neural} to design the VQ loss and set $\beta=0.25$.

To reduce the training complexity and improve the scalability, all neural networks (i.e., $g_v(\cdot), g_e(\cdot), \{g_m^{(t)}(\cdot),g_h^{(t)}(\cdot) \}_{t=1}^T$, $\text{Encoder}(\cdot)$, $\text{Decoder}(\cdot)$, and $g_v^{\text{est}}(\cdot)$) 
 share weights among different nodes and edges.
 After a centralized training stage,  the proposed VQ-MPNN can be implemented in a distributed manner during the testing phase. 
 Each node possesses its own neural network. 
 Even when a node is added or removed from the graph, the proposed VQ-MPNN continues to function, which demonstrates its strong generalizability and scalability.

\section{Simulation Results}

\subsection{Scenario Setting}

\begin{figure}[tbp]
	\centering
	\includegraphics[width=0.73\linewidth]{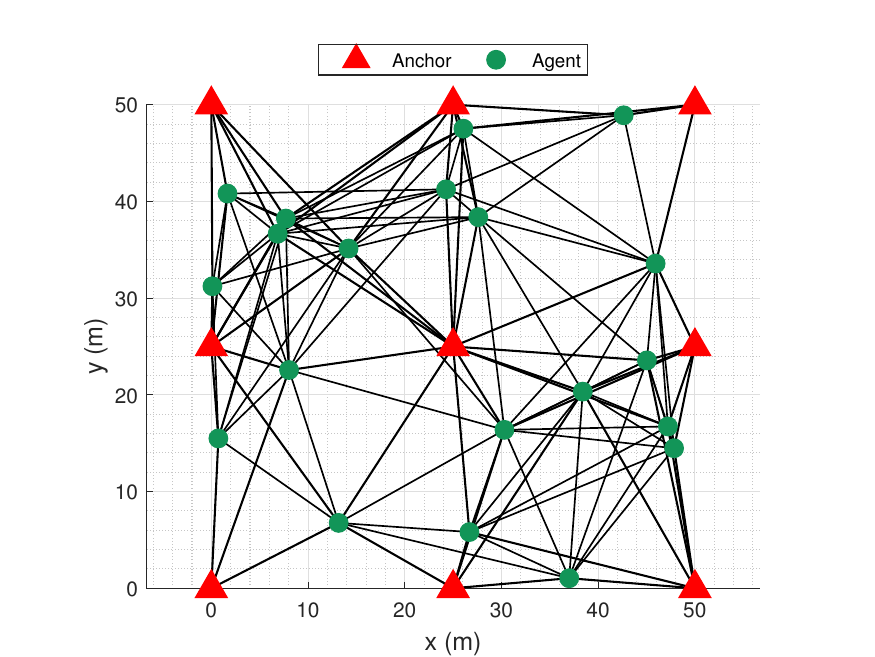}
	\caption{
    An example of a cooperative localization network topology. Red triangles denote the positions of 9 anchors and green circles denote the positions of 20 agents. The communication range is set to 25 $m$.
	} 
	\label{topology}
    \vspace{-6mm}
\end{figure}

We consider a two-dimensional static scenario of area spanning  $[0,50]\, m \times [0,50]\, m$ consisting of 9 anchor nodes and 20 agent nodes.
In particular, the agents are uniformly distributed in this $50\times 50\,(m^2)$ area, and the locations of anchors are set to the four vertices, the midpoints of the sides, and the center of the square area.
The communication range $r$ is set to 25 $m$.
Fig. \ref{topology} illustrates an example of network topology.
The prior distribution of each agent is  $p(\bm{x}_i)\sim \mathcal{N}(\bm{\mu}_i,\bm{\Sigma})$.
The covariance $\bm{\Sigma}$ is given by $\bm{\Sigma} = \text{diag}\{10,10\}$. 
The mean $\bm{\mu}_i$ is randomly drawn from Gaussian distribution $\mathcal{N}(\bm{x}_i,\bm{\Sigma})$.
Following \cite{zhang2023distributed,jin2021exploiting}, we consider two types of measurement noise of distance measurement from node $j$ to $i$: additive white Gaussian noise (AWGN) with zero mean and standard deviation $\sigma$, and range-dependent Gaussian noise with zero mean and a standard deviation of $\sigma\|\bm{x}_j-\bm{x}_i\|_2$.

\subsection{Neural Network Hyperparameters}
The neural network hyperparameters we use for our simulations of VQ-MPNN are given in Table \ref{VQMPNNsetting}.
The dimension of the codeword in the codebook is set to $D=12$ and the dimension of node state is set to $M=16$. 
 We generate 3000 training data samples, i.e., network realizations, to train VQ-MPNN.
	We generate 300 validation data samples to validate
	the performance of neural network. If the best validation loss
	does not change for 30 epochs, the training ends. 
	In the test
	stage, we generate 300 test data samples to test the trained neural
	networks.

\begin{table*}[ht]
    \caption{Comparison of Communication Cost per Node}
    \centering
    \renewcommand{\arraystretch}{1.5} 
    \setlength{\tabcolsep}{6pt} 
    \resizebox{0.99\textwidth}{!}{%
    \begin{tabular}{|c|c|c|c|c|c|c|c|}
        \hline
        \text{Method} & \text{PLBP} & \text{PLBP-$C$ bit} & \text{MPNN} & \text{VQ-MPNN} & \text{SP-ADMM} & \text{NBP} \\ \hline
        \text{Communication Cost (bits)} 
        & $J  N_i  \big(6Q\! + \!H\! + \!4QT\! +\! HT\big)$ 
        & $J N_i  \big(6Q \!+\! H \!+ \!4CT\! + \!HT\big)$
        & $(MQ+H)N_iT$
        & $\big(H + \lceil\log_2 K\rceil\big)N_iT$
        & $(4Q+H)N_iT$ 
        & $(2N_p+H)N_iT$ \\ \hline
    \end{tabular}
    } 
    \label{ComCostcomparison}
    \vspace{-4mm}
\end{table*}

\begin{figure}[htbp]
	\centering  
	\subfigcapskip=0pt 
	\subfigure[RMSE versus communication bits with AWGN model.]{
		\label{Bits_RMSE_1}
		\includegraphics[width=0.8\linewidth]{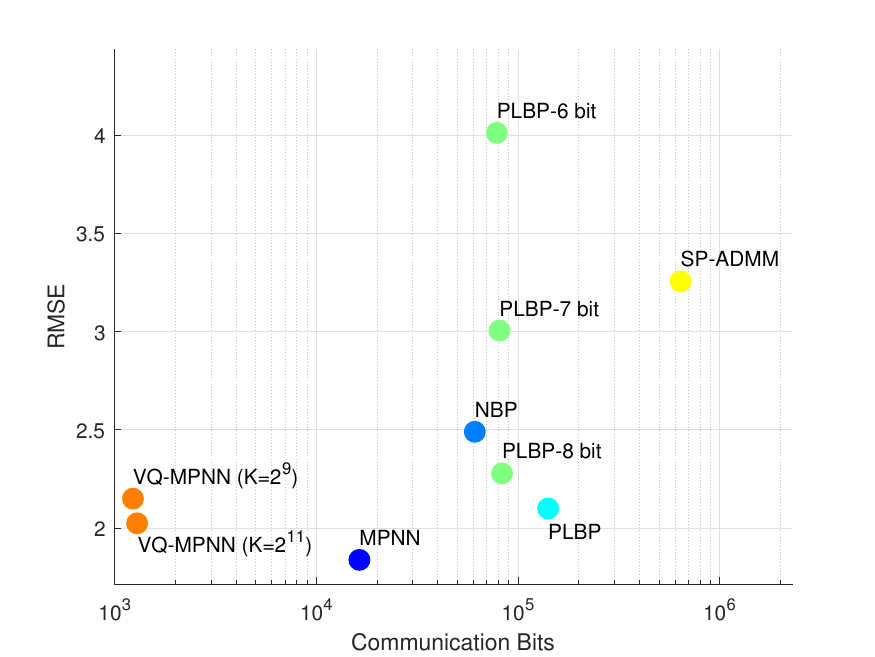}}
	\quad 
	\subfigure[RMSE versus communication bits with range-dependent noise model.]{
		\label{Bits_RMSE_2}
		\includegraphics[width=0.8\linewidth]{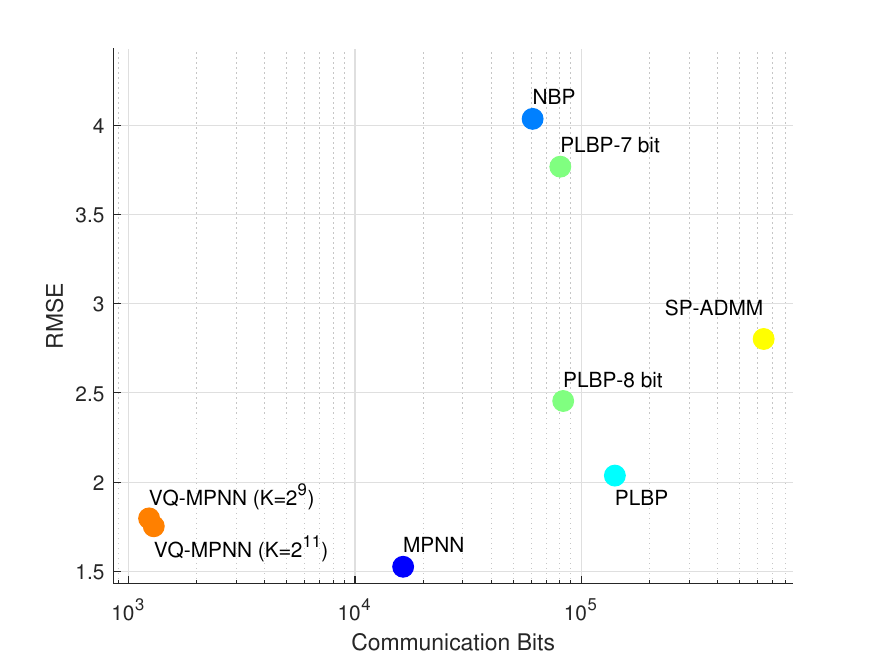}}
	\caption{Communication cost comparison of different algorithms.
    The proposed VQ-MPNN achieves a communication bit count several orders of magnitude smaller than that of other algorithms.} 
    \label{Bits_RMSE}
    \vspace{-5mm}
\end{figure}

\subsection{Performance Metric and Baseline Methods}

We use root mean square error (RMSE) to evaluate performance, which is defined as 
\begin{align}
	\text{RMSE} = \frac{1}{N_{\text{mc}}}\sum_{n=1}^{N_{\text{mc}}}\sqrt{\frac{\sum_{i\in\mathcal{I}_a}\|\hat{\bm{x}}_i(n)-\bm{x}_i(n)\|_2^2}{N_a}},\notag\\
	\forall\, n\in\{1,\ldots,N_{\text{mc}}\},
\end{align}
where $N_{\text{mc}}$ denotes the number of Monte Carlo experiments, $\hat{\bm{x}}_i(n)$ denotes the estimate of node $i$'s  position $\hat{\bm{x}}_i(n)$ in the $n$-th experiment.

The proposed VQ-MPNN is compared with the following baseline methods:
\begin{itemize}
	\item PLBP \cite{garcia2017cooperative}: PLBP is a parametric BP algorithm for cooperative localization.
	PLBP includes double loop iterations: linearization, which calculates linearization parameters, and BP, which calculates beliefs.
	After each linearization of distance measurements, BP is executed on the linearized model.
	Before calculating linearization parameters, means and variances of posteriors, which are represented by a $2\times1$ vector and a $2\times2$ matrix, are exchanged between two nodes to calculate the joint posterior of two nodes.
	Then, linearization parameters are calculated by statistical linear regression with respect to the joint posterior over the two nodes.
	To calculate belief of node $i$, there are $4N_i$ scalars to be exchanged between node $i$ and its neighbors.
    Assuming a fixed packet header overhead per message, the total communication cost per node is given by $J \cdot ((6Q + H)N_i + (4Q+H) N_i T)$ bits, where $Q$ represents the quantization bits, $J$ denotes the number of linearization iterations, $T$ denotes the number of BP iterations and $H$ denotes the communication bits of the packet header.

	\item PLBP-$C$ bit:
	We modify PLBP by using a scalar quantizer to quantize the messages to be exchanged.
	The scalar quantizer is as follows.
	We denote $v$ as the input scalar to the quantizer. 
	We manually set the quantization range $b$. 
	To enforce the scalar within the quantization range $[-b,b]$, we apply a clipping function $\text{clip}(v) = -b + \max(v+b,0)-\max(v-b,0)$. 
	The quantization step size is set as $\delta = \frac{2b}{2^C}$ where $C$ denotes the quantization bits. 
	The quantized scalar is given by
	$
	\hat{v} = \delta \lceil \frac{v}{\delta} \rceil - \frac{\delta}{2}
	$.
    The total communication cost per node is  $J \cdot ((6Q + H)N_i + (4C+H) N_i T)$ bits.

	\item NBP \cite{wymeersch2009cooperative,lien2012comparison}: A nonparametric particle-based BP with beliefs and messages are represented as weighted particles. 
    The total communication cost per node is $(2N_p+H)N_iT$ bits, where $N_p$ denotes the number of particles.
	\item SP-ADMM \cite{zhang2023distributed}: SP-ADMM is an distributed alternating direction method of multipliers (ADMM) algorithm for the nonconvex localization problem.
    In each iteration, each node sends two two-dimensional variables to its neighbors. 
    The total communication cost per node is $(4Q+H)N_iT$ bits.
	\item MPNN \cite{tedeschini2023message}:
	A MPNN is trained to directly estimate agents' positions.
    In each communication round,
    each node sends one $M$-dimensional vector to its
    neighbors.
    The total communication cost for MPNN is $(MQ+H)N_iT$ bits.
\end{itemize}

\subsection{Communication Cost Comparison}
In this subsection, we compare communication cost of all algorithms under two different noise models.
Table \ref{ComCostcomparison} shows the per-node communication costs for all algorithms.
In non-quantized algorithms, scalars are typically  represented as floating-point numbers.
Additionally, the average number of neighbors for each agent node in our scenario is approximately $10$.
Therefore, in the calculations, the quantization bits $Q$ are set to $32$ bits, and the number of neighbors $\mathcal{N}_i$ is set to $10$.
We set $T=3$ for VQ-MPNN and MPNN, NBP, PLBP and PLBP-$C$ bit.
For SP-ADMM, we set $T=400$ to ensure convergence.
Besides, the number of linearization for PLBP and PLBP-$C$ bit is set to 20.
The number of particles for NBP is set to $N_p = 1000$.
The fixed overhead of packet headers per message $H$ is set to 32 bits.

\begin{figure}[tbp]
	\centering
	\includegraphics[width=0.8\linewidth]{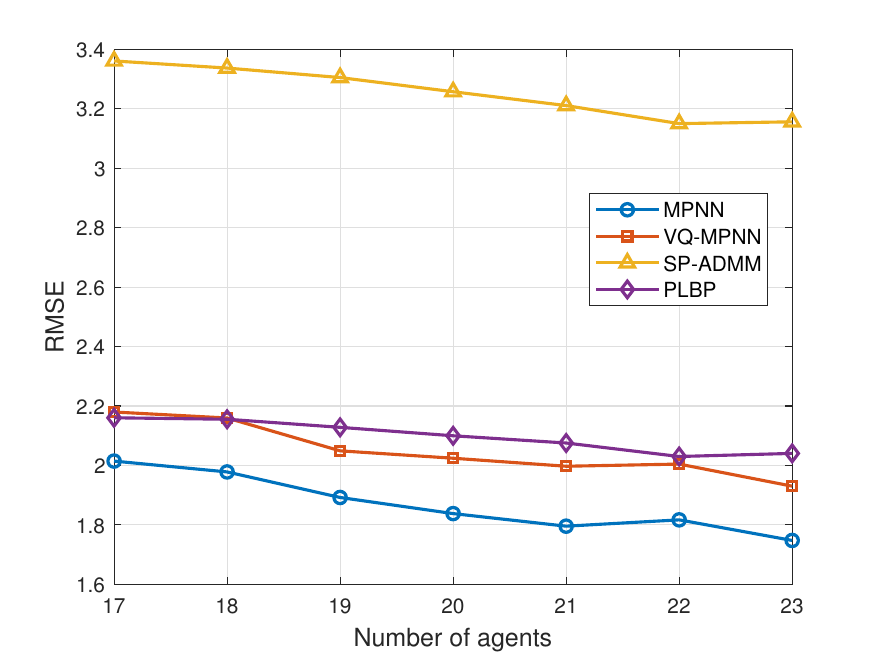}
	\caption{RMSE versus number of agents under AWGN model.
    The well-trained VQ-MPNN is able to generalize to different number of agents.} 
	\label{Agents}
    \vspace{-4mm}
\end{figure}

Based on Table \ref{ComCostcomparison}, Fig. \ref{Bits_RMSE_1} illustrates  the communication bits versus RMSE under AWGN model with a standard deviation of 4.
As shown in Fig. \ref{Bits_RMSE_1}, the MPNN achieves lowest RMSE
 and the proposed VQ-MPNN achieves a second lowest RMSE among all the methods.
The proposed VQ-MPNN and the existing MPNN outperform BP-based methods in wireless networks with loopy
graph topologies, such as the one illustrated in Fig. \ref{topology}. 
This is because the data-driven nature of GNNs is able to refine message accuracy and improve overall positioning performance.
The localization performance of VQ-MPNN improves with a larger codebook, as it provides greater capacity for message representation.
This is because the larger codebook offers greater capacity for message representation.
We observe that PLBP-$C$ bit exhibits a higher RMSE compared to the PLBP due to the quantization error and reducing the quantization bits further increases the RMSE.
Among all the algorithms, VQ-MPNN clearly offers the best tradeoff in localization error and communication efficiency. It requires the least communication bits (at least one order of magnitude smaller compared to other algorithms) while achieving RMSE almost as good as MPNN.

Fig. \ref{Bits_RMSE_2} illustrates  the communication bits versus RMSE under the range dependent noise model with $n_{j\to i}\sim \text{Gauss}(0,0.2\| \bm{x}_j-\bm{x}_i\|_2)$, where $\text{Gauss}()$ denotes the Gaussian distribution.
Similar performance trends in terms of RMSE and communication bits can be observed for all
algorithms under the range dependent noise model. This shows that our approach is robust to the way in which noise is modeled.

\begin{figure}[htbp]
	\centering  
	\subfigcapskip=-4pt
    \subfigure[Loss value versus epoch. The shaded area depicts the mean $\pm$  standard deviation over five random seeds.]{
		\label{Loss_epoch}
\includegraphics[width=0.8\linewidth]{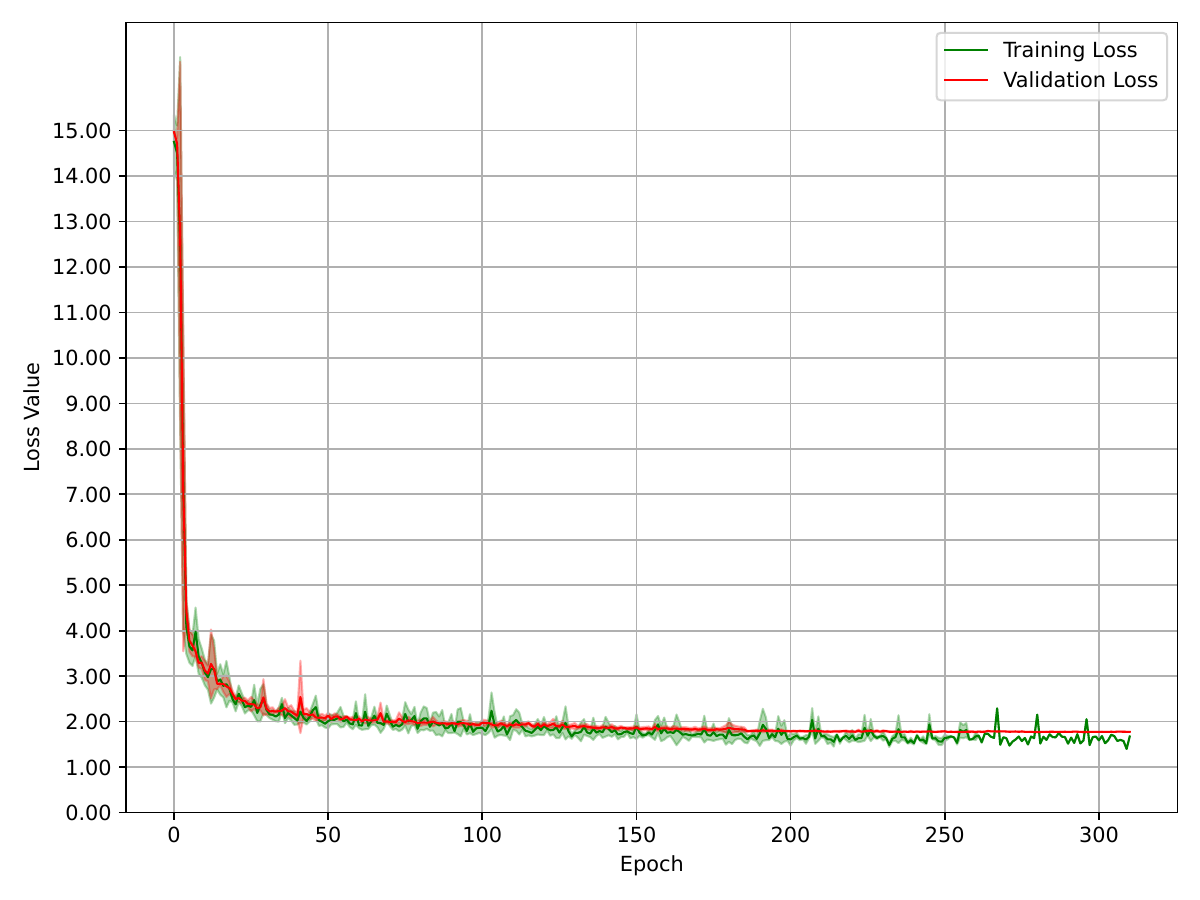}} \\
\vspace{-0.15in}
	\subfigure[Loss value versus training time.]{
		\label{Loss_time}
		\includegraphics[width=0.85\linewidth]{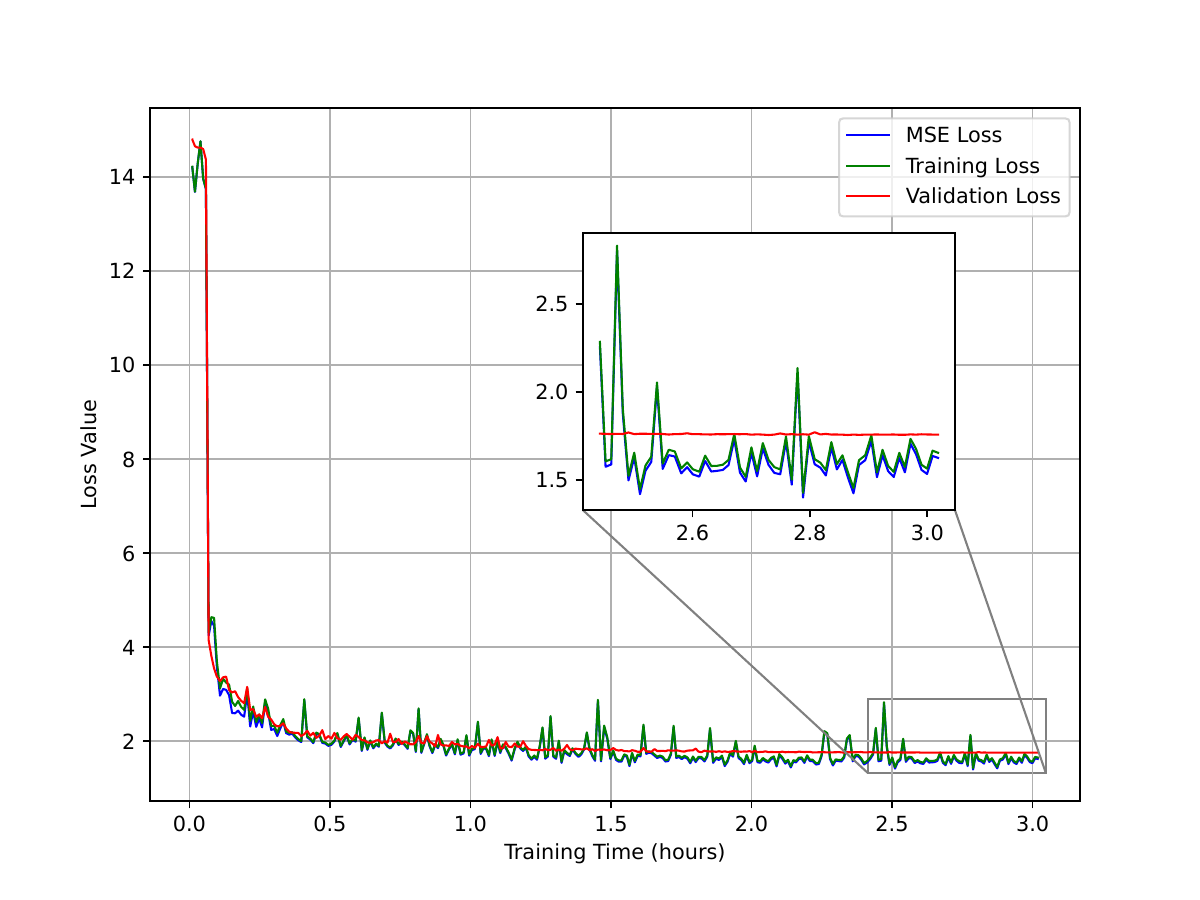}}
    
    \caption{Learning curves for the proposed VQ-MPNN.} 
    \label{learningcurve}
    \vspace{-2mm}
\end{figure}

\subsection{Assessing Generalizability}
In practice, networks are often dynamic, which requires algorithms that can adapt to varying numbers of agents.
This subsection demonstrates that the proposed VQ-MPNN is scalable with respect to varying numbers of agents.
We train VQ-MPNN  and MPNN using a dataset with a fixed number of agents, and subsequently evaluate the well-trained models across varying agent counts.
We train the neural networks with the setup where the number of agents is 20, the number of communication rounds is 3, and the size of the codebook is $2^{10}$.
The number of iterations for SP-ADMM is set to 400.
For PLBP, The number of linearization iterations and BP iterations is set to $20$ and $5$, respectively.
Fig. \ref{Agents} illustrates that RMSE under different number of agents.
We observe that the RMSE decreases with the increase of the number of agents because more neighbors provide more topology information.
The VQ-MPNN trained with  with a fixed number of agents still outperform PLBP and SP-ADMM, indicating that the proposed VQ-MPNN generalizes well across different agent counts.

\subsection{Convergence Behavior}

Fig. \ref{learningcurve} illustrates the training process of the proposed VQ-MPNN with range-dependent noise model of $n_{j\to i}\sim \text{Gauss}(0,0.2\| \bm{x}_j-\bm{x}_i\|_2)$. 
The size of codebook and the number of communication rounds are set to $2^{11}$ and 3.
We train our model on a server equipped with an NVIDIA Tesla V100S PCIe 32GB.
Fig. \ref{Loss_epoch} shows the loss values over epochs for both training and validation, 
The results are obtained by running VQ-MPNN with five different seeds.
We observe that the gap between the training and validation losses remains relatively small throughout the training process, indicating that the model neither overfits nor underfits.
Additionally, the narrow shaded area suggests that the model is both stable and robust.
In Fig. \ref{Loss_time}, we observe that training takes approximately three hours, and the training and validation losses converge to nearly identical values in the final stage. Thus, our method does not require significant computational resources, very large datasets, or time for training. VQ-MPNN provides a quick and efficient data-driven approach to cooperative localization. Additionally, we present the MSE loss in the figure to demonstrate that the gap between the MSE and training losses is minimal, indicating that the VQ loss is very small and our codebook design introduces only a negligible loss due to quantization.

\section{Conclusion}
This paper proposed VQ-MPNN to achieve communication-efficient cooperative localization.
In our approach, prior node position and distance measurements are treated as node and edge features, which are then encoded as node and edge states.
Employing a vector quantized codebook to map the node states to the codewords, each node only need to transmit the indices of the codewords, which significantly enhances communication efficiency. The proposed VQ-MPNN can perform inference on the graph that corresponds to the network topology in a distributed manner.
Simulation results demonstrated that the proposed VQ-MPNN simultaneously achieve low localization error and high communication efficiency.

While we have focused on static networks with independent additive noise, an interesting direction of future work would be to consider mobile networks and possibly correlated noise models for distance measurements. We believe that data-driven methods can strongly outperform traditional approaches in mobile networks with correlated noise, since they are hard to model accurately.

\vspace{-2mm}
\section*{Acknowledgment}
This project was supported in part by the National Science Foundation (NSF) under grant CNS-2212565, the Office of Naval Research (ONR) under grants N00014-21-1-2472 and N00014-23-C-1016, and the Air Force Office of Scientific Research (AFOSR) under grant FA9550-24-1-0083.

\bibliographystyle{IEEEtran}
\bibliography{ref}

\end{document}